\documentclass[conference]{IEEEtran}
\IEEEoverridecommandlockouts

\usepackage{filecontents}
\usepackage{graphicx}
\usepackage{hyperref}
\usepackage{tabularx}
\usepackage{amsmath}
\usepackage{amssymb}
\usepackage{caption}
\usepackage{subcaption}
\usepackage{booktabs}
\usepackage[normalem]{ulem} 

\ifCLASSINFOpdf

\else

\fi

\begin{document}

\title{COLD: Concurrent Loads Disaggregator\\for Non-Intrusive Load Monitoring}

\makeatletter
\newcommand{\linebreakand}{%
  \end{@IEEEauthorhalign}
  \hfill\mbox{}\par
  \mbox{}\hfill\begin{@IEEEauthorhalign}
}
\makeatother

\author{
\IEEEauthorblockN{Ilia Kamyshev\IEEEauthorrefmark{1}\IEEEauthorrefmark{2}, Sahar Moghimian Hoosh\IEEEauthorrefmark{1}\IEEEauthorrefmark{2}, Dmitrii Kriukov\IEEEauthorrefmark{1}\IEEEauthorrefmark{3}, Elena Gryazina\IEEEauthorrefmark{1}, Henni Ouerdane\IEEEauthorrefmark{1}}
\IEEEauthorblockA{\IEEEauthorrefmark{1}\textit{Skoltech}, Moscow, Russia}
\IEEEauthorblockA{\IEEEauthorrefmark{2}\textit{Monisensa Development LLC.}, Moscow, Russia}
\IEEEauthorblockA{\IEEEauthorrefmark{3}\textit{Artificial Intelligence Research Institute}, Moscow, Russia}
\IEEEauthorblockA{Emails: \{Ilia.Kamyshev, Sahar.Moghimian, Dmitrii.Kriukov, E.Gryazina, H.Ouerdane\}@skoltech.ru}
}

\maketitle

\begin{abstract}
The global effort toward renewable energy and the electrification of energy-intensive sectors have significantly increased the demand for electricity, making energy efficiency a critical focus. Non-intrusive load monitoring (NILM) enables detailed analyses of household electricity usage by disaggregating the total power consumption into individual appliance-level data. In this paper, we propose COLD (Concurrent Loads Disaggregator), a transformer-based model specifically designed to address the challenges of disaggregating high-frequency data with multiple simultaneously working devices. COLD supports up to 42 devices and accurately handles scenarios with up to 11 concurrent loads, achieving 95\% load identification accuracy and 82\% disaggregation performance on the test data. In addition, we introduce a new fully labeled high-frequency NILM dataset for load disaggregation derived from the UK-DALE 16 kHz dataset. Finally, we analyze the decline in NILM model performance as the number of concurrent loads increases. 

\end{abstract}


\IEEEpeerreviewmaketitle

\section{Introduction}
\label{sec:org841953a}

Global energy markets are shifting from the over-exploitation of fossil fuels to renewable energy sources, which drives the electrification of different energy-intensive sectors worldwide \cite{IEA_2024}. The shift towards electrification is expected to nearly double global electricity demand by 2050, while fossil fuel demand is projected to plateau by 2030 \cite{IEA_2024}. Without improving energy efficiency, renewable energy and electrification alone cannot mitigate the environmental impact of energy-intensive economies \cite{IEA_2024}. Given the increased demand for electricity, one of the most promising techniques to improve energy efficiency on the end user side is Non-intrusive load monitoring (NILM) which provides detailed insights into electricity usage by disaggregating the total end-user's power consumption into the consumption of individual appliances (see Fig.~\ref{fig:overview}). By identifying specific appliance activities, NILM empowers users to make informed decisions about their energy usage, and reduce electricity consumption by 10–15\% \cite{anderson2012event}.

\begin{figure}[t!]
\centering
 \includegraphics[width=\columnwidth]{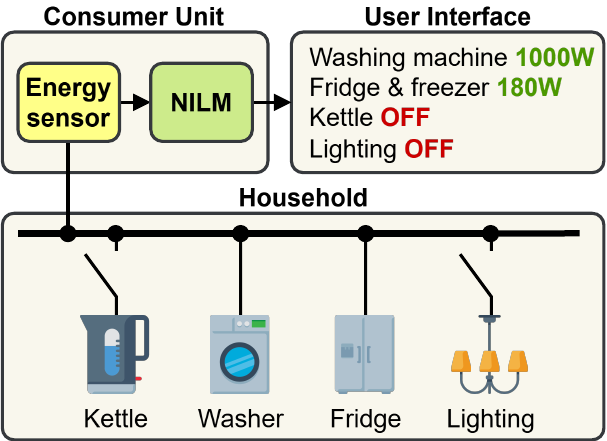}
\caption{Overview of the NILM technology}
\label{fig:overview} 
\end{figure}


NILM systems analyze the current and voltage entering a household to detect operating appliances, determine their status, and their energy consumption. Each appliance exhibits a distinct energy usage pattern, commonly referred to as its “signature” \cite{zoha2012non}. Appliance signatures can be extracted at two sampling rates: high-frequency (in kHz ranges) and low-frequency (1 Hz or less). High-frequency data, which is the focus of this paper, captures microscopic features such as harmonic compositions resulting from circuit non-linearities. These features provide an additional layer of detailed information which can significantly enhance the accuracy of energy disaggregation algorithms, particularly in scenarios where a large number of appliances operate at the same time \cite{kamyshev2023edframe}. 

Over the years, with advances in deep learning and the increased availability of data, deep neural networks have become the primary focus of NILM research \cite{deeplr}. Convolutional neural network (CNN) models have been widely used in the literature as, e.g., in \cite{de2024st, zhang2020non}, whereas novel networks based on long short-term memory (LSTM) for energy disaggregation were proposed in  \cite{kaselimi2020context, xia2020non}. A number of studies have implemented 2D convolutions \cite{schirmer2021double, fryze}. The approach proposed in \cite{fryze} transforms high-frequency current signals into an image-like representation, which is then used as input to a CNN for multi-label classification. Moreover, transfer learning and attention mechanisms have recently emerged as promising techniques to enhance NILM performance. For instance, BERT4NILM \cite{bert4nilm}, a self-attention mechanism and bidirectional transformer model for low-frequency data, successfully outperformed CNN, LSTM.

It must be noted that the most important aspect of a NILM model is its disaggregation ability. By this, we mean how effectively a model can distinguish multiple “simultaneously” operating appliances (load identification) and accurately estimate their respective power shares (energy estimation per each appliance). While models often perform well with only two or three electrical devices, their performance in complex scenarios with a larger number of overlapping appliances remains underexplored. NILM studies such as \cite{semisup, tp}, typically select three or five commonly used appliances from publicly available datasets for their experiments. However, an exploratory analysis of well-known datasets such as UK-DALE \cite{ukdale} indicates that practical scenarios may involve more complex combinations. 

In this regard, we analyzed the statistics from the UK-DALE dataset. Figure \ref{fig:ukdale} illustrates the distribution of concurrently active appliances in House 1 from UK-DALE. 
\begin{figure}[b]
        \centering
        \includegraphics[width=\columnwidth]{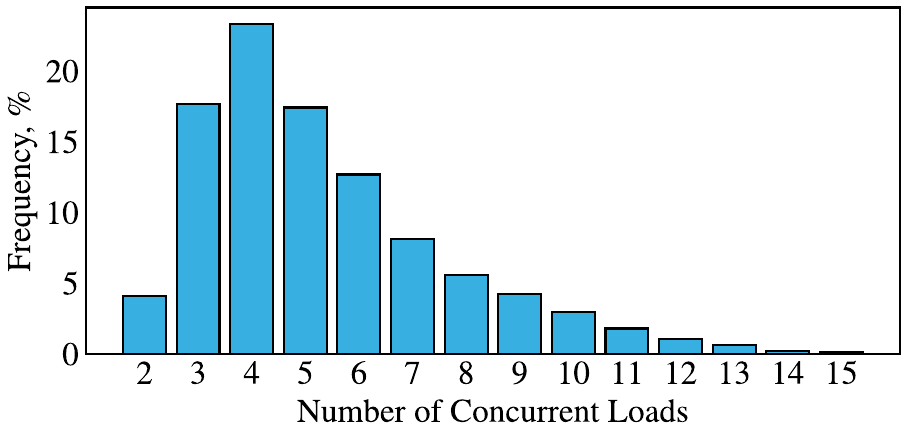}
        \caption{\label{fig:ukdale}Distribution of the number of simultaneously active devices in the House 1 from the UK-DALE dataset.} 
\end{figure}
While four appliances are mostly operating at the same time, higher numbers (five or more) are also observed. This observation highlights that disaggregating a larger number of concurrent loads remains an open challenge. Although a few studies, such as \cite{schirmer2021double, fryze, faustine2020unet}, have touched on this issue, to the best of our knowledge, no existing research has systematically investigated the sensitivity of NILM model performance to the number of concurrently operating appliances.


\subsection{Our contributions}
To resolve the issue of disaggregating a large number of concurrent loads, we propose a novel transformer-based architecture named COLD (Concurrent Loads Disaggregator). Unlike most existing deep learning models for NILM, COLD is a multi-output/multi-label and multi-task model designed to perform load identification and load disaggregation for high-frequency datasets. COLD supports the recognition of up to 42 different devices and handles scenarios with a maximum of 11 devices working simultaneously while maintaining high disaggregation accuracy (at least 60\%). The model was evaluated across two previously unseen real-world households and achieves superior load identification and disaggregation scores, 95\% and 82\% respectively, compared to other approaches e.g., \cite{schirmer2021double,fryze}. In addition to COLD, other contributions of this work are as follows:

\begin{itemize}
\item We obtained a fully labeled high-frequency NILM dataset from the famous UK-DALE dataset. The dataset is ready-to-use and contains 85k training steady-state voltage-current signals, 5k validation signals, and 10k test signals.
\item We studied the performance decline of NILM models with the number of concurrent loads growing and provided three explanations for this phenomenon.
\item The source code for COLD and the obtained dataset are available in the GitHub repository:

\href{https://github.com/arx7ti/cold-nilm}{https://github.com/arx7ti/cold-nilm}
\end{itemize}

This paper is organized as follows: Section \ref{sec:ukdale} is devoted to the dataset preparation process. In section \ref{sec:model}, we present the COLD architecture and its implementation. Section \ref{sec:results} covers the results of the model evaluation and our comparison with existing NILM models. Section \ref{sec:discussion} provides a comprehensive discussion of the findings. Finally, Section \ref{sec:conclusion} concludes the paper and suggests future research directions.

\section{Preparation of UK-DALE dataset}
\label{sec:ukdale}

For real-time load identification and disaggregation, the COLD model operates on high-frequency NILM data. For model inference, we used UK-DALE dataset \cite{ukdale}, which includes aggregated 16 kHz voltage-current readings and submetered active power measurements with an average latency of 6 seconds. This means a 6-second current waveform corresponds to one root mean square (RMS) value per appliance. We downloaded 17,750 one-hour recordings from House 1 of the dataset which spans around 3 TB on SSD. We used House 1 since it contains the majority of the recordings, and comprises 52 different devices. To reduce the dataset size, we extracted six-second steady-state segments from the one-hour recordings. Note, we excluded segments after on/off events due to possible misalignment between high-frequency aggregated and low-frequency submetered data. To extract those segments, we first identified change points using a z-score event detector applied to active power within a six-second window (84-second window size, 10 z-score threshold). The window was reset after each detected event. The total power consumption signal was then divided into 18-second frames between successive change points. Steady-state segments were selected as frames where the standard deviation remained stable (below z-score threshold 0.5).

We put 7 devices (LED printer, ADSL router, child's lamp, iPad charger, iron, office lamp 3, soldering iron) into the category "other" since their presence in combination with other devices was negligibly small. Finally, the obtained dataset resulted in 100,000 fully labeled voltage-current waveforms sampled at 16 kHz, with 42 device labels (including category "other") and corresponding active power shares. We also labeled devices that consume less than 10 W at a time as "other." The resulting high-frequency dataset can be used for both multi-label classification and multi-output regression tasks of energy disaggregation. See Fig.~\ref{fig:spec}(a) for an example of a training signal. Last, we split the obtained dataset into train, validation, and test subsets, each containing 85k, 5k, and 10k signals, respectively.

\section{Concurrent Loads Disaggregator (COLD)}
\label{sec:model}

\begin{figure}[t!]
        \centering   
       \includegraphics[width=\columnwidth]{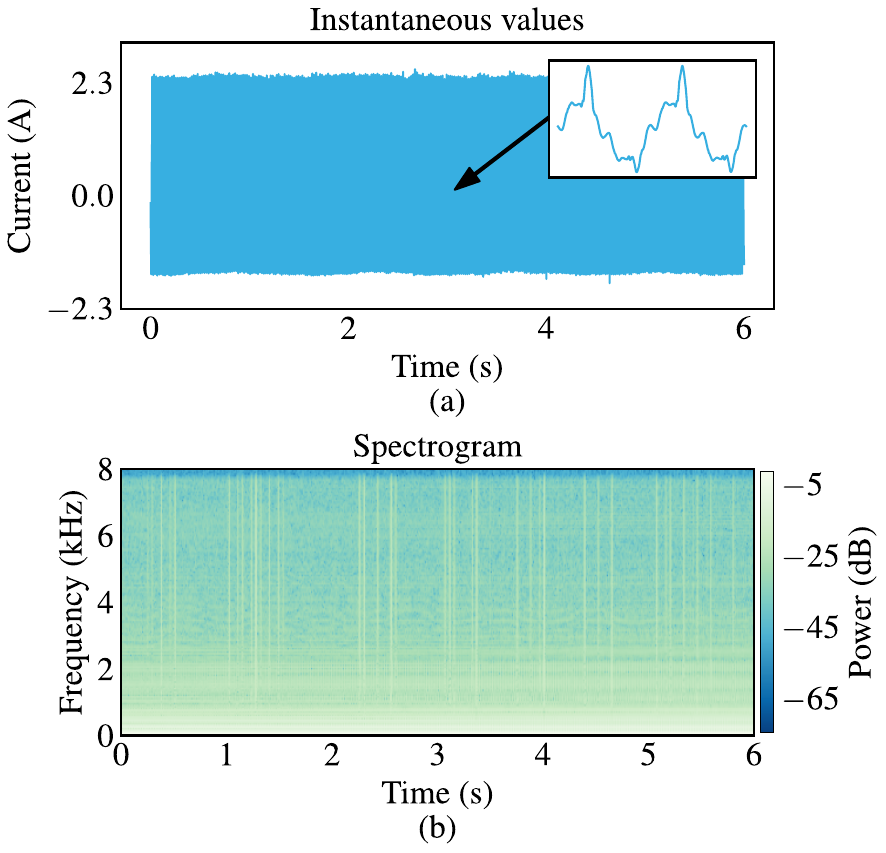}
        \caption{\label{fig:spec}One of the steady-state segments extracted from the House 1 of the UK-DALE dataset (a) and its spectrogram (b).} 
\end{figure} 

We propose the model named COLD (Concurrent Loads Disaggregator) which has a transformer-based architecture, depicted in Fig.~\ref{fig:architecture}. COLD is an end-to-end NILM model designed to solve a multi-output regression task i.e., energy disaggregation, and multi-label classification task, i.e., load identification.

\begin{figure*}[t!]
        \centering
       \includegraphics[width=0.88\textwidth]{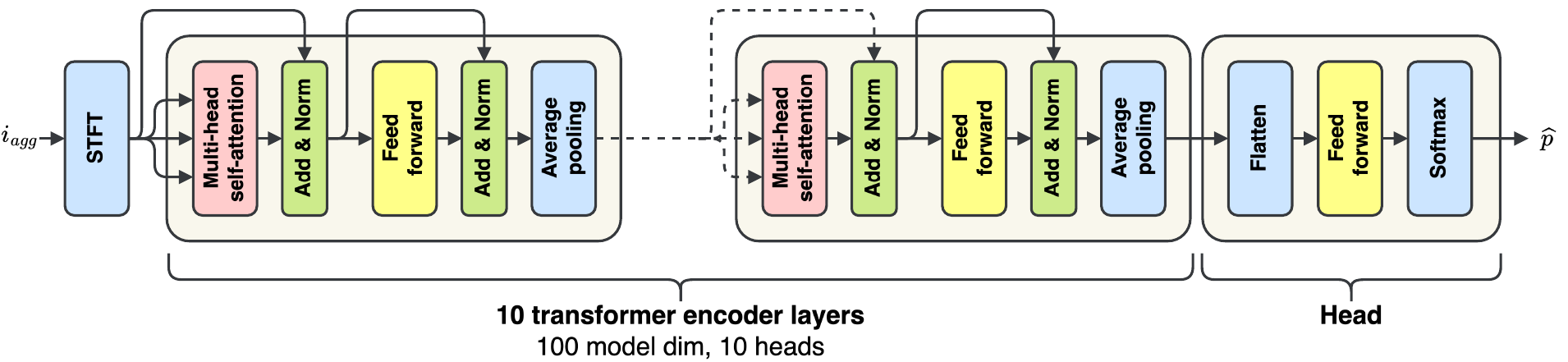}        \caption{\label{fig:architecture}COLD architecture layout.} 
\end{figure*}

The transformer architecture is the state-of-the-art model originally developed for natural language processing due to its ability to capture long-range dependencies and contextual relationships using self-attention mechanisms \cite{vaswani2017attention}. The choice of this transformer architecture for NILM is motivated by its analogy to the machine language translation tasks. Indeed, both tasks requires mapping one representation (the aggregate signal or the source language) to another representation (the disaggregated shares or the target language).  

\subsection{Features extraction}

COLD takes spectrograms as input features. From each aggregated current signal we extract spectrograms with the use of short time Fourier transform (STFT). We set the window size for Fourier transform to 20 ms, the time between each successive window to 10 ms. Then the spectrogram is converted to the decibel scale as shown in Fig.~\ref{fig:spec}(b).

Finally, we apply z-score normalization by subtracting the global mean -19.29 dB and the global standard deviation 7.21 dB over each spectrogram. Thus, the entire training set has zero mean and unit variance. We denote $X=\text{STFT}(i)$ as a matrix of input features, where $i$ is an aggregated instantaneous values of the current. We also set the number of harmonics to 50, including the offset.

\subsection{Transformer encoder layers}

The architecture of COLD consists of ten successive and lightweight transformer encoder layers. Each layer has a base dimensionality ($d_{model}$) of 100 neurons, and 400 neurons in the projection layer of the feed-forward network. We add an adaptive average pooling at the end of each transformer layer to reduce the size of the spectrogram along the time axis. Thus, the initial spectrogram of size $600\times 50$ will become $1\times 100$ after the first transformer layer and will retain its size till the head layer. The reason why we apply such rapid compression along the time frames is that the current signal is in a steady state. Hence, each frame of a spectrogram does not deviate far from its mean. One should notice that this is recommended only for the given dataset. Other datasets might require sequential reduction of a spectrogram across the time axis. Below, we briefly explain the essence of the multi-head self-attention mechanism.

\subsection{Multi-head self-attention}

To learn the time dependence in a spectrogram, a multi-head self-attention (MHSA) is used. This mechanism is the key element for the transformer encoder layer. It is called multi-head as it splits the input features by several non-overlapping subsets (heads) and learns their representations independently. The term "self" implies that it does not require separate query, key, and value, but learns them from an input vector of features. Finally, it performs \emph{scaled dot-product attention} in parallel $n_{head}$ times, then concatenates the results and projects onto another learnable space:

\begin{equation}\label{eqn:mhsa}
\begin{split}
    \text{MHSA}(X)&=\text{Concat}\{\text{Head}_i:i\in[1,n_{head}]\}P^T\\
    &\text{where}\ \text{Head}_i=\text{Attention}(Q_i, K_i, V_i)
\end{split}
\end{equation}
where $Q_i=XW_{Q,i}^T$, $K_i=XW_{Q,i}^T$, $V_i=XW_{V,i}^T$ are query, key and values, respectively. Matrices $W_{Q,i}$, $W_{K,i}$ and $W_{V,i}$ are projection matrices for each head $i$. The matrix $P$ is an output projection matrix.

The scaled dot-product attention is the following function:
\begin{equation}
    \text{Attention}(Q_i,K_i,V_i)=\text{Softmax}\left(\frac{Q_iK_i^T}{d_h}\right)V_i
\end{equation}

The scaling by parameter $\frac{1}{d_h}$ is needed to prevent softmax's output values from its saturated region i.e., to mitigate vanishing gradient problem. It is recommended to set $d_h=d_{model}/n_{head}=100/10=10$.


\subsection{Head of COLD}


In the final layer, also referred to as the head, we use a single-layer feed-forward network with a softmax activation function. Prior to the feed-forward layer, we apply matrix flattening to ensure consistency within the architecture. Flattening is necessary when the transformer's output features span more than one time frame. The width of the feed-forward network's is set to 100 neurons. The softmax ensures that the output power shares sum to one, satisfying the \textit{energy conservation law} i.e., the sum of disaggregated power shares cannot be more or less than the aggregated power.

To retrieve the absolute power shares, the relative power shares can be multiplied by the aggregated power. Once final power shares are obtained, the head layer performs thresholding i.e., suppressing power shares below 10W to exclude models' noise. Finally, each processed power share is being divided by the sum of all processed power shares. 

\subsection{Loss function}

To train COLD, we used the binary cross-entropy loss:
\begin{equation}
    \mathcal{L}(p, \hat{p})=-\frac{1}{n_d}\sum_{d} [p_d\log (\hat{p}_d)+(1-p_d)\log(1-\hat{p}_d)]
\end{equation}
where vectors $p_d$ and $\hat{p}_d$ are actual and disaggregated power shares of device $d$, respectively. Both vectors are normalized such that the sums of their elements equal to 1. The total number of devices is denoted as $n_d$.

\subsection{Evaluation metrics}

To evaluate the quality of disaggregation of concurrent loads, we employ two metrics for energy estimation: the modified F1-score (MF) \cite{kim2011unsupervised} and the total energy correctly assigned (TECA)\cite{makonin2015nonintrusive}. Additionally, we use the conventional F1-score metric (F1) for evaluating the load identification. To calculate MF, the following two counters need to be computed:

\begin{equation}
\begin{split}
    \text{ATP}=\sum 1\left(\frac{|\hat{p}_d-p_d|}{p_d}\geq \delta\,\land\,p_d > 0\right),\\
    \text{ITP}=\sum 1\left(\frac{|\hat{p}_d-p_d|}{p_d}<\delta\,\land\,p_d > 0\right),
\end{split}
\end{equation}

\noindent where $1(\cdot)$ is an indicator function that returns one if the condition is true, and zero otherwise. The threshold $\delta$ was set to $0.2$. Accurate true positives (ATP) and inaccurate true positives (ITP) represent the correctly and incorrectly estimated active power of a correctly identified devices, respectively. ATP and ITP are essential for evaluating both the model's load identification capability and its energy estimation performance. False positives (FP) and false negatives (FN) count cases where a load was falsely identified as being on or off. Thus, MF can be computed as:

\begin{equation}
    \text{MF}=\frac{\text{ATP}}{\text{ATP}+\text{ITP}+0.5(\text{FP}+\text{FN})}
\end{equation}

The conventional F1-score requires true positives (TP) which can be computed as $TP=ATP+ITP$, then:

\begin{equation}
    \text{F1}=\frac{\text{TP}}{\text{TP}+0.5(\text{FP}+\text{FN})}
\end{equation}

As mentioned earlier, TECA is another metric used to evaluate energy estimation, and it is defined as follows:

\begin{equation}
    \text{TECA}=1-\frac{\sum_d{|\hat{p}_d-p_d|}}{2\sum_d p_d}.
\end{equation}

\section{Results}
\label{sec:results}

The COLD model was trained using the Adam \cite{adam} optimizer with cosine annealing learning rate scheduling \cite{scheduling}. We set the minimal learning rate to $10^{-7}$ and the maximum to $10^{-3}$ with a period of 30 epochs. As shown in Fig.~\ref{fig:train-val}, the model converges approximately after 600 epochs on a 85k training dataset. For experiments, we used the workstation with Intel i7-14700K CPU, and two NVIDIA RTX 4070 Ti GPUs.

\begin{figure}[b!]
\centering
 \includegraphics[width=\columnwidth]{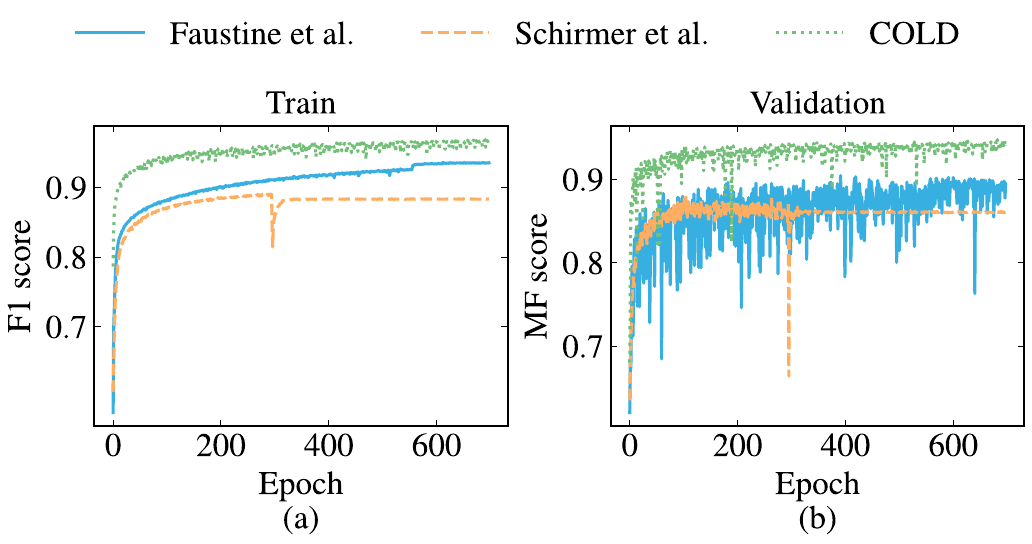}
\caption{\label{fig:train-val}F1 and MF score curves over 700 training epochs.}
\end{figure}

To evaluate the generalization ability of our model, we conducted power disaggregation for 10,000 test signals, and computed F1, MF and TECA metrics. As a result, COLD achieves a 95\% F1-score in load identification, along with 93\% TECA and 82\% MF in load disaggregation, see Table~\ref{tab:scores}. The distributions of the load identification and disaggregation scores over different appliances are given in Figs.~ \ref{fig:apps}(a) and \ref{fig:apps}(b), respectively. According to Fig.~\ref{fig:ukdale}, four devices are active most of the time: data logger PC, fridge, boiler, and devices from category "other". Their respective F1 (MF) scores are: 99\% (93\%), 98\% (90\%), 92\% (67\%), and 100\% (87\%).

\begin{figure*}[tbp]
\centering
 \includegraphics[width=\textwidth]{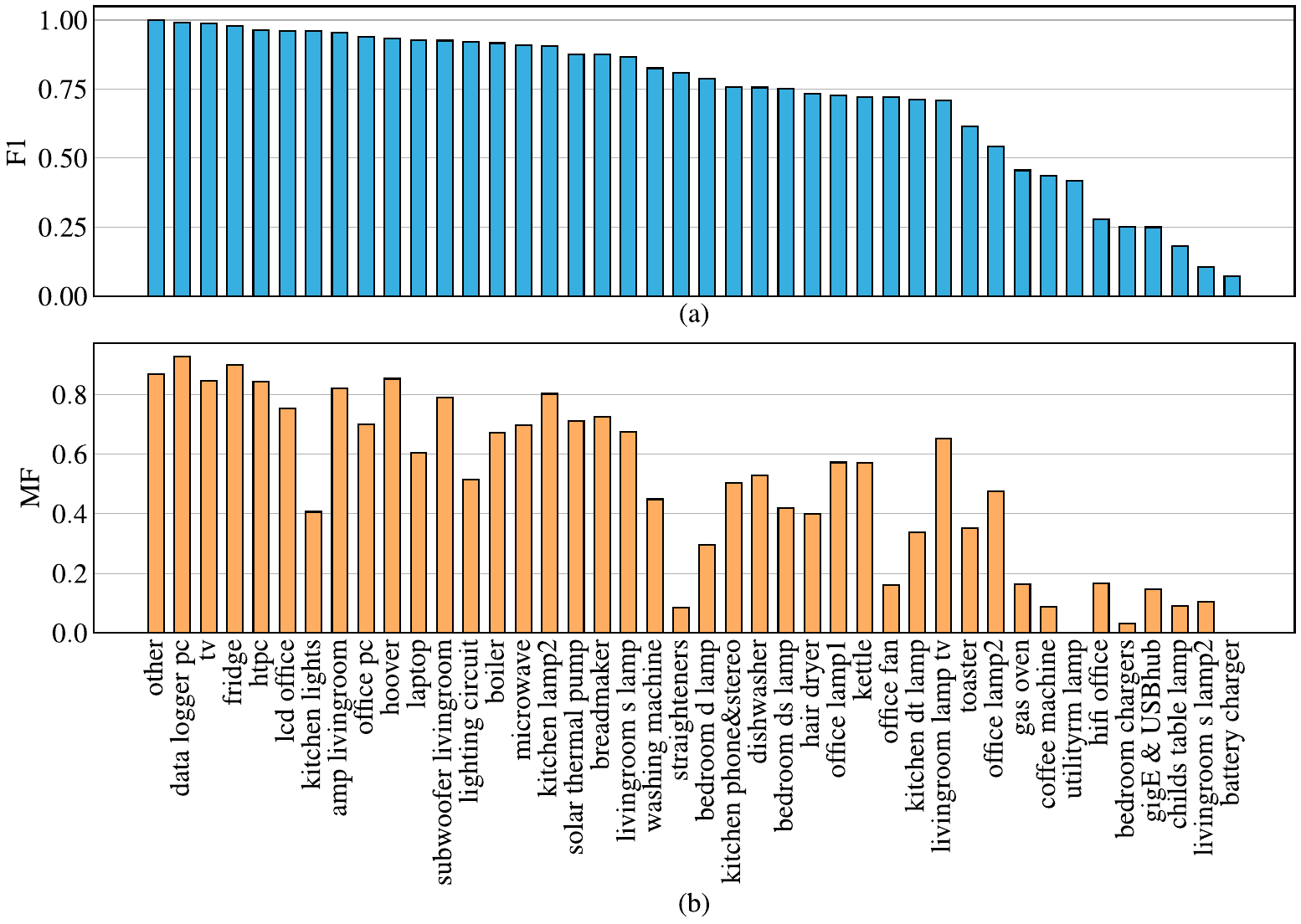}
\caption{\label{fig:apps}F1 (a) and MF (b) scores computed for each individual appliance from the House 1 of UK-DALE dataset.}
\end{figure*}

\begin{table}[t!]
\caption{\label{tab:scores}Comparison of high-frequency NILM models.}
\centering
\begin{tabularx}{\linewidth}{>{\hsize=1.5\hsize\linewidth=\hsize}X>
{\hsize=0.7\hsize\linewidth=\hsize}X>{\hsize=0.5\hsize\linewidth=\hsize}X>{\hsize=0.5\hsize\linewidth=\hsize}X>{\hsize=0.5\hsize\linewidth=\hsize}X}
\toprule
Model & Features & F1 & TECA & MF \\
\midrule
Schirmer et al. \cite{schirmer2021double} & DFIA & 0.87 & 0.85 & 0.58 \\

Faustine et al. \cite{fryze} & DD & 0.90 & 0.90 & 0.72 \\

\textbf{COLD} (ours) & STFT & \textbf{0.95} & \textbf{0.93} & \textbf{0.82} \\
\bottomrule
\end{tabularx}
\end{table}

Next, we evaluated both load identification and disaggregation performance across varying numbers of simultaneous loads. To the best of our knowledge, this work is the first in the field that shows the effectiveness of the NILM model versus the number of concurrent loads. In Fig.~\ref{fig:concurrent}, we show that all three metrics F1, TECA and MF are decreasing monotonically with a growing number of simultaneous loads. However, load identification performance is less prone to it. On the other hand, unlike F1 and TECA, the MF drastically decreases with the number of concurrent loads growing. MF shows a faster decrease since the average disaggregation error for COLD is 14 W which might result in an increased number of ITP for devices with comparable power consumption e.g., chargers. By setting minimal MF to 60\%, COLD can identify up to 11 concurrent loads with high degree of accuracy. Threshold 60\% implies that more than a half of test cases were handled successfully by COLD.

We also trained two notable models for high-frequency data proposed by Schirmer et al. \cite{schirmer2021double} and Faustine et al. \cite{fryze}, on the so-called double Fourier integral analysis (DFIA) and decomposed-distance (DD) features. These features were extracted from the training signals of the obtained dataset. Results show that COLD achieves both superior load identification and load disaggregation performance. That is, 8\% and 5\% better than Schirmer et al. and Faustine et al. in load identification, respectively. Regarding load disaggregation, COLD is from 8\% to 24\% (TECA, MF) more efficient than Schirmer et al., and from 3\% to 10\% more accurate than Faustine et al. 

\begin{figure}[t!]
\centering
 \includegraphics[width=\columnwidth]{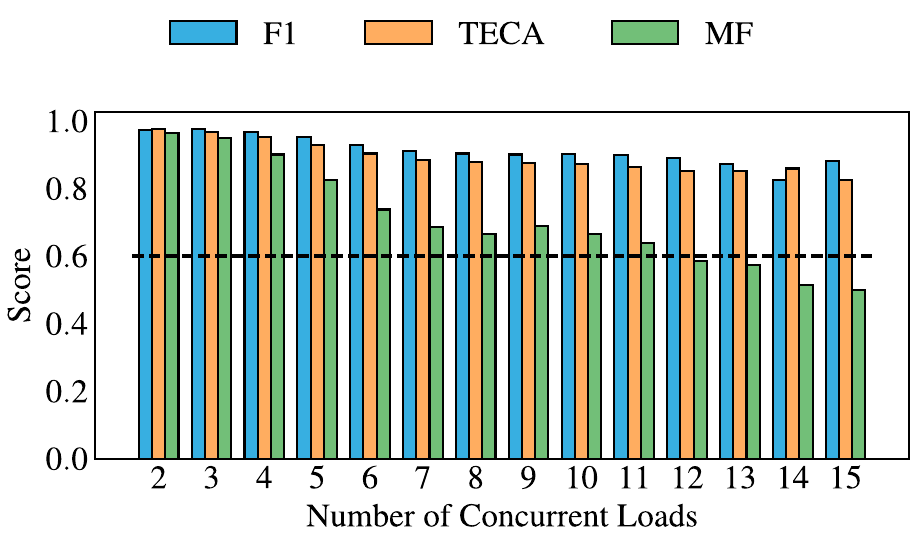}
\caption{\label{fig:concurrent}F1, TECA and MF computed for COLD for different numbers of simultaneously operating appliances. Dashed line indicates 60\% score.}
\end{figure}

\section{Discussion}
\label{sec:discussion}

Our hypotheses regarding the decline in NILM model performance as the number of concurrent loads increases can be explained as follows:

\begin{itemize}
   \item \textbf{A strong correlation between signatures}.
   Some appliances can be built similarly but have different applications. This can result in either highly similar or identical signatures. For example, both boiler and kettle are designed to heat water but for different purposes; hence, their high-frequency power signatures might be the same in terms of form factor (sinusoidal waveforms).

   \item \textbf{Shift to the "blind" zone of the model}.
   As the number of concurrent loads increases, the relative power share of each appliance in the total consumption decreases. The effect is stronger when more powerful appliances are turned on. Hence, appliances with smaller power shares may fall below the detection threshold of the NILM model (10 W for COLD), entering entering a so-called "blind" zone where their power shares assigned to the category "other".
   
   \item \textbf{Linear combination of signatures}.  
   Multiple appliances operating simultaneously increase the likelihood of devices whose total consumption (linear combination) resembles the signature of another device. For example, a washing machine consists of at least three components that can be found as individual devices in many households: an AC/DC converter, a heating element, and a water pump. 
   
\end{itemize}

All the points stated above are under the assumption that the model is trained fairly and the data is balanced i.e., no appliance is more likely to be active than another. Data imbalance might introduce bias of scores in a favor of most frequently occurring devices.

\section{Conclusion}
\label{sec:conclusion}

For the first time in the field of NILM, we studied the decline in NILM model performance as the number of simultaneous loads increases. We proposed a novel transformer-based architecture, COLD, designed to improve both load identification and disaggregation accuracy. COLD achieves state-of-the-art performance on high-frequency data: 95\% for load identification and 82\% for load disaggregation. Moreover, we processed UK-DALE 16 kHz dataset to create a ready-to-use, fully labeled high-frequency NILM dataset for load identification and disaggregation tasks. The dataset includes 85k training signals, 5k validation and 10k test signals, respectively. In future work, we plan to design a loss function that takes into account the average power of each device to mitigate the problem of disaggregating low-power devices.

\ifCLASSOPTIONcaptionsoff
  \newpage
\fi

\bibliographystyle{IEEEtran}
\bibliography{confpaper}
\end{document}